\pdfoutput=1

\documentclass[11pt]{article}

\usepackage[final]{acl}

\usepackage{times}
\usepackage{latexsym}

\usepackage[T1]{fontenc}

\usepackage[utf8]{inputenc}

\usepackage{microtype}

\usepackage{inconsolata}

\usepackage{graphicx}

\usepackage{threeparttable}
\usepackage{amsmath}
\usepackage{amsfonts}
\usepackage{booktabs}
\usepackage{enumitem}   
\usepackage{adjustbox}
\usepackage{bm}

\title{Eta-WavLM: Efficient Speaker Identity Removal in Self-Supervised Speech Representations Using a Simple Linear Equation}

\author{
  \textbf{Giuseppe Ruggiero\textsuperscript{1,2}},
  \textbf{Matteo Testa\textsuperscript{2}},
  \textbf{Jurgen Van de Walle\textsuperscript{2}},
  \textbf{Luigi Di Caro\textsuperscript{1}}
\\
  \textsuperscript{1}Università degli studi di Torino, Turin, Italy \\
  \textsuperscript{2}Cerence Inc, Turin Italy
\\
    \{giuseppe.ruggiero, luigi.dicaro\}@unito.it \\ \{matteo.testa, jurgen.vandewalle\}@cerence.com
}

\begin{document}
\maketitle

\begin{abstract}
Self-supervised learning (SSL) has reduced the reliance on expensive labeling in speech technologies by learning meaningful representations from unannotated data. Since most SSL-based downstream tasks prioritize content information in speech, ideal representations should disentangle content from unwanted variations like speaker characteristics in the SSL representations. However, removing speaker information often degrades other speech components, and existing methods either fail to fully disentangle speaker identity or require resource-intensive models. In this paper, we propose a novel disentanglement method that linearly decomposes SSL representations into speaker-specific and speaker-independent components, effectively generating speaker disentangled representations. Comprehensive experiments show that our approach achieves speaker independence and as such, when applied to content-driven tasks such as voice conversion, our representations yield significant improvements over state-of-the-art methods.\footnote{Audio samples for the voice conversion system are available at: \url{https://giuseppe-ruggiero.github.io/eta-wavlm-vc-demo/}}
\end{abstract}

\section{Introduction}
\label{sec:introduction}
In recent years, speech-related tasks such as automatic speech recognition (ASR), text-to-speech (TTS), voice conversion (VC), and speech-to-speech translation (S2S) have made significant advancements, achieving near-human performance in several domains. However, these high-performing systems often rely on large quantities of high-quality labeled data, which is both resource-intensive and time-consuming to obtain, limiting speech technologies scalability across languages, domains, and applications. 

To address this challenge, researchers have increasingly focused on techniques that leverage vast amounts of unlabeled data for model training. Among these, self-supervised learning (SSL) has emerged as a transformative paradigm, enabling models to learn latent representations from raw input data without the need for explicit labels. In the speech domain, the core concept of SSL is to pretrain a speech representation network on large-scale unannotated corpora, with the objective of capturing and encoding meaningful speech structures and information~\citep{contentvec}. SSL models such as Wav2Vec2~\citep{wav2vec2}, HuBERT~\citep{hubert}, and WavLM~\citep{wavlm} have shown great success in extracting robust and versatile features directly from speech waveforms. These SSL representations can then be exploited for downstream tasks using only a limited amount of labeled data~\citep{nansy}.

SSL representations encode diverse speech attributes, including linguistic content, speaker identity, emotions, and background conditions, making them versatile but often task-agnostic. For example, a good representation for tasks like VC or TTS should be rich in content but contain minimal to no speaker identity \citep{a-comparative-study, contentvec}, while speaker classification or verification prioritizes speaker information. Consequently, disentangling speaker and non-speaker information in SSL representations is a critical aspect to improve task-specific performance \citep{softvc, ace-vc, repcodec, base-tts, enhancing-polyglot-voices}, though it remains highly challenging \citep{contentvec, basetts-starting-point}. To this end, SSL representations are often quantized to derive pseudo-text from speech utterances, with k-means clustering being a widely used technique due to its simplicity and unsupervised nature \citep{hubert, speech-resynthesis, per-utt-std}. However, this often also compromises linguistic content and prosody \citep{basetts-starting-point, enhancing-polyglot-voices}. 

To address this issue, alternative disentanglement strategies have been proposed. These include strategies based on simple perturbation techniques applied to the input waveform \citep{nansy, ace-vc}, utterance-level standardization of representations \citep{per-utt-std, vectok}, neural models and training conditions designed to extract content-related-only features from SSL representations \citep{contentvec, softvc, repcodec}, and the incorporation of specific model components, training strategies, or loss functions to achieve  disentanglement online during the training phase in tasks like VC or TTS \citep{basetts-starting-point, base-tts}. Although these methods preserve content better than k-means, many still struggle to achieve a high level of disentanglement \citep{enhancing-polyglot-voices} or require the implementation of complex and resource-intensive strategies.

In this paper, we propose a novel and general approach for disentangling the speaker identity from SSL representations without requiring complex training strategies, loss functions, fine-tuning, or even quantization. We show that SSL representations can be \textit{linearly} decomposed into speaker-dependent $\mathbf{d}$ and speaker-independent $\bm{\eta}$ components, which we will refer to as \textit{eta representations}. This means that, if $\mathbf{d}$ is known, the speaker-independent \textit{eta} representation can be easily obtained by solving a linear inverse problem. 

Our main contributions are as follows: 1) We introduce an efficient disentanglement strategy for generating speaker-independent SSL representations by solving a simple linear equation; 2) We demonstrate that our method actually generates speaker-independent representations, reducing speaker accuracy in a speaker-related classification task by nearly 30\% compared to standard SSL representations; 3) We show that the features derived from our approach enhance the performance of a task-specific VC model. Specifically, our approach improves target speaker identity, linguistic content preservation, and overall system quality. These findings align with the hypotheses of prior work \citep{contentvec, basetts-starting-point}, indicating that effectively addressing speaker disentanglement can yield significant performance improvements in content-related speech tasks.

\section{Method}
\label{sec:method}
\begin{figure}[t]
  \centering
  \includegraphics[width=\columnwidth]{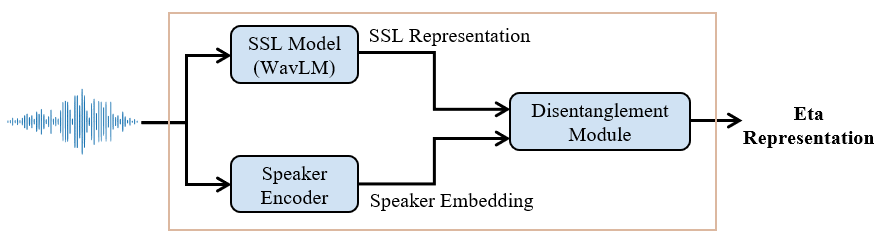}
  \caption{High-level overview of the proposed approach.}
  \label{fig:architecture}
\end{figure}

The proposed approach can be considered an extension of an SSL model, implemented as an offline module designed to extract disentangled \textit{eta} representations. As illustrated in Figure \ref{fig:architecture}, our method consists of three key components: an \textit{SSL model} that extracts an SSL representation from a raw waveform, a \textit{speaker encoder} that generates a speaker embedding from the same waveform, and a \textit{disentanglement module} which derives a speaker-independent \textit{eta} representation from the input SSL representation, conditioned on the speaker embedding. In this work, both the SSL and the speaker encoder modules are off-the-shelf pre-trained models that are not further trained or fine-tuned. Our main contribution lies in the implementation of the disentanglement module.
\subsection{Problem Definition}
\label{subsec:problem-definition}
The primary concept underlying the disentanglement module is to decompose an SSL representation $\mathbf{s}$, into speaker-dependent $\mathbf{d}$ and speaker-independent $\bm{\eta}$ components. For a given data point, $\mathbf{s}$ and $\mathbf{d}$ can be easily obtained using a pre-trained SSL model and a pre-trained speaker encoder, respectively. Thus, $\mathbf{s}$ can be expressed as a function of the known $\mathbf{d}$ along with an additional unknown term $\bm{\eta}$, which encapsulates all the information not inferable from $\mathbf{d}$. For simplicity, we assume an additive relationship which can be described as:
\begin{equation}
    \mathbf{s} = f(\mathbf{d}) + \bm{\eta}
\end{equation}
Ideally, $\bm{\eta}$ should include linguistic, prosodic, and information from the environment (e.g. recording conditions), provided that $\mathbf{d}$ effectively represents speaker characteristics. The importance of selecting an appropriate speaker encoder for extracting $\mathbf{d}$ in this context will be discussed in Section \ref{subsec:speaker-encoder-and-pca}. Consequently, the speaker-independent component $\bm{\eta}$ can be computed as:
\begin{equation}
    \bm{\eta} = \mathbf{s} - f(\mathbf{d})
\end{equation}
In the next section, we will discuss how to model the function $f()$.

\subsection{Computation of Latent Basis and Bias}
\label{subsec:computation-of-latent-basis-and-bias}
Based on the hypothesis that large embedding spaces tend to linearize complex non-linear relationships~\citep{linear-word-analogies, speaker-and-phonetic-info-in-ssl}, we approximate $f()$ using a linear model. Consider a multi-speaker dataset composed of $U$ utterances of raw speech. Let us denote a generic utterance as $\mathbf{u}_i$, its speaker embedding extracted by a pre-trained speaker encoder $\mathcal{E}$ as $\mathbf{e}_i \in \mathbb{R}^V$ with $i \in [1, U]$, and its SSL representation extracted by a pre-trained SSL model $\mathcal{S}$ as $\mathbf{S}_i = [\mathbf{s}_1, \cdots, \mathbf{s}_M]^T$, where $\mathbf{s}_m \in \mathbb{R}^Q$ represents the $m$-th frame, and $M$ is the sequence length. Since $M$ can be large, we randomly subsample $L$ frames from each utterance, creating a fixed-length representation $\mathbf{S}_i \in \mathbb{R}^{L \times Q}$. Consequently, the entire dataset's SSL representation, obtained by stacking all the $\mathbf{S}_i$, is given by $\mathbf{S} \in \mathbb{R}^{N \times Q}$, where $N = U \times L$ for simplicity. 

To align $\mathbf{e}$ with the sequence length of $\mathbf{S}$, we leverage the fact that the speaker embedding captures speaker-level information, which is assumed to remain constant across all frames of an utterance. Based on this, we expand $\mathbf{e}$ by replicating it $L$ times along the frame axis, resulting in $\mathbf{E}_i \in \mathbb{R}^{V \times L}$. Consequently, the entire dataset's embedding representation, obtained by stacking all the $\mathbf{E}_i$, is given by $\mathbf{E} \in \mathbb{R}^{V \times N}$. In addition, since $V$ can be large, we apply Principal Component Analysis (PCA) to reduce its dimension to $P < V$, thus obtaining $\mathbf{D} \in \mathbb{R}^{P \times N}$. This reduction helps remove redundancy and retains only the most informative components. We will show the importance of this step in Section \ref{subsec:speaker-encoder-and-pca}. 

Given $\mathbf{S} \in \mathbb{R}^{N \times Q}$ and $\mathbf{D} \in \mathbb{R}^{P \times N}$, we can model their relationship as: 
\begin{align}
    \mathbf{S} = \mathbf{D}^T \mathbf{A} + \mathbf{1}_N \mathbf{b}^T
\end{align}
where $\mathbf{A} \in \mathbb{R}^{P \times Q}$ and $\mathbf{b} \in \mathbb{R}^{Q \times 1}$ are learnable parameters. For simplicity, we can rewrite it as:
\begin{align}
    \mathbf{S} = \mathbf{\tilde{D}}^T \mathbf{\tilde{A}}
\end{align}
where $\mathbf{\tilde{D}}^T = \begin{bmatrix} \mathbf{D}^T & \mathbf{1} \end{bmatrix} $ and $\mathbf{\tilde{A}}^T = \begin{bmatrix} \mathbf{A}^T & \mathbf{b} \end{bmatrix}$. Then, the optimization problem we want to solve is given by:
\begin{align}
    \mathbf{\tilde{A}}^* = \arg\min_{\mathbf{\tilde{A}}} || \mathbf{S} - \mathbf{\tilde{D}}^T \mathbf{\tilde{A}} ||_F
\end{align}
which can be solved through the pseudo-inverse as:
\begin{align}
    \mathbf{\tilde{A}}^* = (\mathbf{\tilde{D}}^T \mathbf{\tilde{D}})^{-1} \mathbf{\tilde{D}}^T \mathbf{S}
\end{align}
where $\mathbf{\tilde{A}}^*{^T} = \begin{bmatrix} \mathbf{A}^*{^T} & \mathbf{b}^* \end{bmatrix}$. From now on, $\mathbf{A}^*$ and $\mathbf{b}^*$ will be referred to as latent basis and bias.

At this stage, the function $f()$ has been learned, marking the completion of the first step. With $\mathbf{A}^*$ and $\mathbf{b}^*$ known, the disentanglement module is now able to generate \textit{eta} representations.

\subsection{Creation of Eta Representations}
\label{subsec:creation-of-eta-features}
During the inference phase, the proposed system (Figure \ref{fig:architecture}) generates speaker-independent \textit{eta} representations directly from raw waveforms. Given an utterance $\mathbf{u}'$, first the pre-trained SSL model $\mathcal{S}$ extracts an SSL representation $\mathbf{S} \in \mathbb{R}^{K \times Q}$: 
\begin{align} 
    \mathbf{S} = \mathcal{S}(\mathbf{u}'; \mathbf{W_{\mathcal{S}}}) 
    \label{eq:ssl-eq} 
\end{align} 
where $\mathbf{W_{\mathcal{S}}}$ represents the frozen parameters of the SSL model, and $K$ is the sequence length. Next, the pre-trained speaker encoder $\mathcal{E}$ generates a speaker embedding $\mathbf{e} \in \mathbb{R}^{V \times 1}$: 
\begin{align} 
    \mathbf{e} = \mathcal{E}(\mathbf{u}'; \mathbf{W_{\mathcal{E}}}) 
    \label{eq:speaker-encoder-eq} 
\end{align} 
where $\mathbf{W_{\mathcal{E}}}$ represents the frozen parameters of the speaker encoder. To reduce the dimensionality of $\mathbf{e}$, PCA is applied, producing $\mathbf{d} \in \mathbb{R}^{P \times 1}$: 
\begin{align} 
    \mathbf{d} = \mathcal{PCA}(\mathbf{e}; \mathbf{C_{\mathcal{PCA}}}) 
    \label{eq:pca-eq} 
\end{align} 
where $\mathbf{C_{\mathcal{PCA}}}$ denotes the matrix of principal components obtained during the PCA process executed in the first step (Section \ref{subsec:computation-of-latent-basis-and-bias}). Finally, the disentanglement module $\mathcal{H}$ extracts a speaker-independent \textit{eta} representation $\bm{\eta} \in \mathbb{R}^{K \times Q}$: 
\begin{align} 
    \bm{\eta} = \mathcal{H}(\mathbf{S}; \mathbf{d}, \mathbf{A}^*, \mathbf{b}^*) 
    \label{eq:eta-generic-eq} 
\end{align} 
where $\mathbf{A}^*$ and $\mathbf{b}^*$ are the latent basis and bias obtained at the end of the first step (Section \ref{subsec:computation-of-latent-basis-and-bias}), and $\mathcal{H}()$ is implemented as: 
\begin{align} 
    \mathcal{H}(\mathbf{S}) = \mathbf{S} - \mathbf{1}_K(\mathbf{d}^T \mathbf{A}^* + \mathbf{b}^*) 
    \label{eq:eta-eq} 
\end{align} 
In this work, we specifically chose WavLM as the SSL model $\mathcal{S}$ for our experiments. Therefore, we will refer to the SSL representations $\mathbf{S}$ as \textit{WavLM representations} and the output of our system $\bm{\eta}$ as \textit{Eta-WavLM representations}.

\section{Experiments}
\label{sec:experiments}
To evaluate the effectiveness of our proposed approach, we selected a speaker-related and a content-related task. The primary objective of the first experiment is to determine whether the \textit{eta} representations extracted by our method exhibit minimal or no speaker-specific characteristics, thereby confirming the achievement of the desired disentanglement. The goal of the second experiment is to assess whether the disentangled representations provides benefits in real-world tasks such as VC, where maintaining linguistic content and achieving high similarity to the target speaker's voice are essential. For all of our experiments, we employed the following setup: 1) \textbf{Framework and hardware}: We ran all experiments on a Linux machine with a single NVIDIA GeForce RTX 3090 GPU with 24 GB of RAM; 2) \textbf{Dataset}: We used the full training set of the multi-speaker LibriSpeech \citep{librispeech} dataset for computing $\mathbf{A}^*$ and $\mathbf{b}^*$, as described in Section \ref{subsec:computation-of-latent-basis-and-bias}. LibriSpeech consists of nearly 1,000 hours of English speech and is openly available under the CC BY 4.0 license; 3) \textbf{SSL model}: As mentioned in Section \ref{subsec:creation-of-eta-features}, we adopted the state-of-the-art WavLM \citep{wavlm} as the pre-trained SSL model $\mathcal{S}$. We used the official WavLM-Large\footnote{\url{https://huggingface.co/microsoft/wavlm-large}} model released under the CC BY-SA 3.0 license and, following \citep{how-can-a-bad-teacher, wav2vec-u, enhancing-polyglot-voices}, we employed the output of the 15th transformer layer as the representation $\mathbf{S}$. Accordingly, we set $Q = 1024$ in Sections \ref{subsec:computation-of-latent-basis-and-bias} and \ref{subsec:creation-of-eta-features}, corresponding to the dimensionality of the WavLM-Large output vectors. In addition, we set $L = 100$ in Section \ref{subsec:computation-of-latent-basis-and-bias}; 4) \textbf{Speaker encoder}: We chose the state-of-the-art ECAPA-TDNN \citep{ecapa-tdnn} as the pre-trained speaker encoder model~$\mathcal{E}$. ECAPA-TDNN extracts speaker embeddings from input speech by leveraging channel attention, propagation, and aggregation mechanisms to produce robust and discriminative speaker representations $\mathbf{d}$. We used a publicly available ECAPA-TDNN model\footnote{\url{https://huggingface.co/speechbrain/spkrec-ecapa-voxceleb}} pre-trained by SpeechBrain \citep{speechbrain} and released under the Apache-2.0 license. Accordingly, we set $V = 192$ in Sections \ref{subsec:computation-of-latent-basis-and-bias} and \ref{subsec:creation-of-eta-features}, corresponding to the dimensionality of the embeddings extracted by the model. The choice of ECAPA-TDNN as the speaker encoder is justified in Section \ref{subsec:speaker-encoder-and-pca}; 5) \textbf{Dimensionality reduction}: We used PCA\footnote{\url{https://scikit-learn.org/1.6/modules/generated/sklearn.decomposition.PCA.html}} to reduce $V$ to $P$, as described in Sections \ref{subsec:computation-of-latent-basis-and-bias} and \ref{subsec:creation-of-eta-features}. We set $P = 128$, and the motivation is discussed in Section \ref{subsec:speaker-encoder-and-pca}.

\subsection{Speaker-Related Classification Task}
\label{subsec:speaker-related-task}
\begin{figure*}[t]
    \centering
    \includegraphics[width=0.90\linewidth]{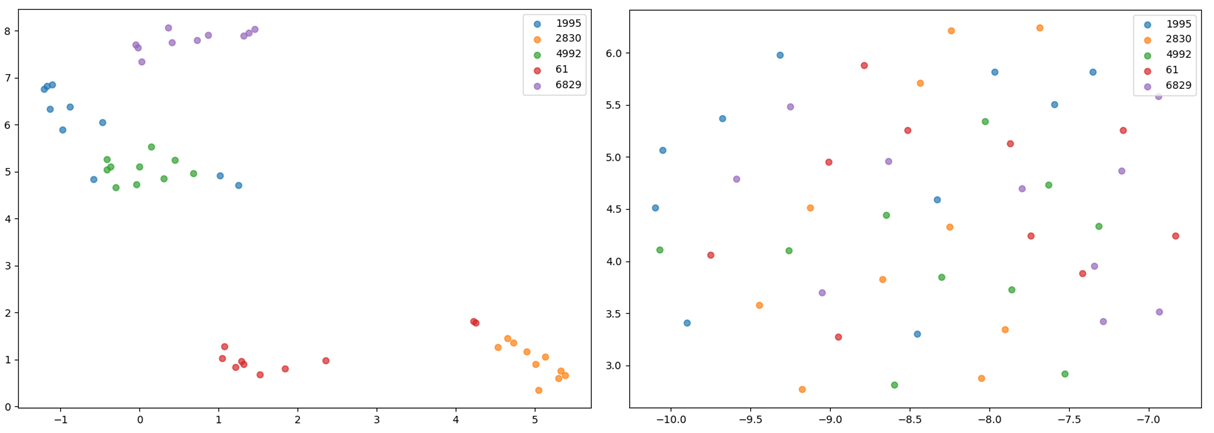}
    \caption{UMAP projections of the WavLM (a) and Eta-WavLM (b) representations extracted from 10 utterances of 5 speakers (with ids 1995, 2830, 4992, 61, 6829) from the LibriSpeech test-clean set.}
    \label{fig:umap-visualization}
\end{figure*}
To evaluate whether the proposed approach effectively reduces speaker information in the WavLM representations, thereby creating speaker-independent Eta-WavLMs, we designed a speaker classification task. Intuitively, since this is a speaker-related task, a model can only perform well if the input representations retain a significant amount of speaker-specific information. Conversely, if the input representations are speaker-independent, the model will struggle to achieve high classification accuracy. Thus, our hypothesis is that our representations will perform worse on the speaker classification task than the original WavLMs, which are known to encode speaker-specific characteristics. To test this hypothesis, we randomly selected 10 speakers from the LibriSpeech test-clean set, resulting in a total of 1285 utterances. Then, for each utterance, we computed the corresponding WavLM representation $\mathbf{S}$ as described in Equation \ref{eq:ssl-eq} and the Eta-WavLM representation $\bm{\eta}$ as described in Equation \ref{eq:eta-eq}. We trained and evaluated a multi-class support vector machine (SVM) classifier on both representation sets using a 5-fold cross-validation setup, recording the classification accuracy for each fold. In addition, we reported the mean and the standard deviation across the 5 folds. We chose SVM for its simplicity and well-known robustness in handling high-dimensional feature spaces and small datasets. The results are shown in Table \ref{tab:spk-acc-wavlm-vs-eta}. 
\begin{table}[th]
    \begin{center}
        \begin{adjustbox}{max width=\columnwidth}
            \begin{tabular}{l|ccccc|c}
                \toprule
                \multicolumn{1}{l|}{} & \bf FOLD1 & \bf FOLD2 & \bf FOLD3 & \bf FOLD4 & \bf FOLD5 & \bf MEAN ± STD \\
                \midrule
                WavLM&   83.46& 82.33& 80.85& 83.30& 81.55& 82.30 ± 0.01 \\
                Eta-WavLM&   53.82& 55.14& 58.77& 53.94& 56.96& \bf 55.73 ± 0.01 \\
                \bottomrule
            \end{tabular}
       \end{adjustbox}
       \caption{Classification accuracy results (\%) for WavLM and Eta-WavLM across the 5 folds of cross-validation (ACC \(\downarrow\)). Lower accuracy indicates better performance, as it reflects reduced speaker-related information.}
        \label{tab:spk-acc-wavlm-vs-eta}
    \end{center}
\end{table}

As expected, the Eta-WavLM representations achieve significantly lower classification accuracy compared to the original WavLM ones (paired t-test yielded a \textit{T-Statistic} of 18.41 and a \textit{p-value} of $5.12 \times 10^{-5}$, rejecting the null hypothesis $p < 0.05$). This accuracy reduction provides clear evidence that our approach is effective in reducing speaker-specific information from the standard WavLM representations.

To further validate our approach, we visualized the WavLM and Eta-WavLM representations. We randomly selected 5 speakers from the LibriSpeech test-clean set and extracted 10 utterances per speaker. For each utterance, we computed the WavLM and Eta-WavLM representations and projected them onto a two-dimensional space using UMAP \citep{umap} (see Appendix \ref{app:pacmap} for a complementary visualization using PaCMAP). As shown in Figure \ref{fig:umap-visualization} (a), the UMAP projection of the WavLM representations cluster to regions corresponding to the individual speakers, suggesting the presence of strong speaker-specific information. In contrast, the projection of the Eta-WavLM representations in Figure \ref{fig:umap-visualization} (b) does not show a discernible cluster of speakers, indicating that our transformation effectively minimizes speaker-specific information. These visualizations reinforce the quantitative results from the speaker classification task by providing an intuitive and qualitative demonstration of the speaker-independence of the Eta-WavLM representations. The absence of speaker-specific clusters in the Eta-WavLM projection aligns with the significantly lower speaker classification accuracy observed in Table \ref{tab:spk-acc-wavlm-vs-eta}, further reinforcing the conclusion that our approach successfully disentangles speaker-related information.

\subsection{Voice Conversion Task}
\label{subsec:voice-conversion}
Despite providing good insights into the reduction of speaker-related information in our representations, the speaker classification task does not assess whether this reduction affects other critical components, such as linguistic content. To evaluate this, we designed a content-related VC task, where the preservation of linguistic content and the accurate representation of the target speaker’s identity are both essential for creating a high-quality conversion system. This dual requirement makes VC an ideal framework for evaluating whether our approach removes speaker-specific components while preserving other essential features. Our hypothesis is that the proposed Eta-WavLM representations will improve VC performance compared to both the original WavLM representations and other state-of-the-art disentanglement methods. 

\subsubsection{Model Architecture}
\label{subsubsec:model-architecture}
For this experiment, we selected the state-of-the-art Any-to-One VC system proposed in \citep{softvc, enhancing-polyglot-voices}, as it achieves impressive levels of linguistic content preservation, target speaker identity similarity, and high-quality speech generation. The architecture consists of a content encoder, an acoustic model, and a vocoder. The content encoder extracts speech representations from a raw waveform of any speaker, the acoustic model converts these representations into a mel spectrogram of the target speaker, and the vocoder synthesizes the resulting mel spectrogram into a speech waveform of the target speaker. 

In this system, we focus on the content encoder, as its role is to extract SSL representations from speech. This makes it the ideal component for incorporating our Eta-WavLM representations and comparing them with other baseline approaches. In contrast, we left the acoustic model unchanged from the implementation in \citep{softvc, enhancing-polyglot-voices} and we trained it from scratch following the original configurations. Further details on its architecture can be found in Appendix \ref{app:vc-am}. For the vocoder, we opted for the multi-speaker Vocos \citep{vocos}, known for its ability to produce high-quality speech outputs. We used the official pre-trained model\footnote{\url{https://huggingface.co/charactr/vocos-mel-24khz}} released under the MIT license.

\subsubsection{Baseline Approaches}
\label{subsubsec:vc-baselines}
We evaluated our Eta-WavLM approach against several baselines, including the direct use of the WavLM model as in \citep{enhancing-polyglot-voices}, as well as four prominent disentanglement strategies from the literature: perturbation, per-utterance standardization, soft unit creation, and Vector Quantization (VQ). Specifically, for perturbation, we implemented the disentanglement strategy outlined in \citep{nansy}, which is based on information perturbation applied to the input speech before the WavLM model. For per-utterance standardization, we employed the utterance-level standardization method described in \citep{per-utt-std} on the WavLM representations. For soft unit creation, we followed the training procedure outlined in \citep{softvc} to derive soft speech units from the WavLM representations. For the VQ strategy, we trained the RepCodec model \citep{repcodec} following the official instructions\footnote{\url{https://github.com/mct10/RepCodec}}, substituting HuBERT with WavLM. This comparison yielded 6 distinct content encoders: one based on the unmodified WavLM representations and five derived from the application of the different disentanglement strategies (including our approach). Each content encoder produces either continuous or discrete representations, depending on the specific disentanglement method applied or whether the output of the WavLM model is used directly without further refinement.
\begin{table*}[t]
\caption{Objective and subjective evaluation of the VC task. Results (\%) in terms of intelligibility (W/PER \(\downarrow\)), target speaker similarity (T-SSIM \(\uparrow\)), source speaker similarity (S-SSIM \(\downarrow\)), and overall quality (MOS \(\uparrow\)) with 95\% confidence intervals for the proposed Eta-WavLM and the baseline methods. WavLM was used as the SSL model across all approaches.}
\label{tab:objective-and-subjective-evalution}
\begin{center}
    \resizebox{\textwidth}{!}{
        \begin{tabular}{l|ccccc|ccccc}
            \toprule
            \multicolumn{1}{l|}{} & \multicolumn{5}{c|}{\bf LJSpeech} & \multicolumn{5}{c}{\bf Elliot Miller}  \\
            \midrule
            \multicolumn{1}{l|}{} & \bf WER & \bf PER & \bf T-SSIM & \bf S-SSIM & \bf MOS \
             & \bf WER & \bf PER & \bf T-SSIM  & \bf S-SSIM & \bf MOS \\
            \midrule
            Ground truth&   3.22&   5.47&   -&  -& 3.85 ± 0.04&  3.22&   5.47&   -&  -& 3.85 ± 0.04\\
            \midrule
            Perturbation \citep{nansy}&  6.29&  7.32&   91.69&  50.64& 3.45 ± 0.06&  10.76& 8.43&   87.41&  52.67& 3.13 ± 0.07\\
            Utterance std \citep{per-utt-std}&  4.13&  7.32&   90.34&  51.58& 3.80 ± 0.04&  5.16& 6.68&   85.91&  55.87& 3.41 ± 0.06\\
            Soft \citep{softvc}&  4.82&  5.94&   91.81&  50.11& 3.84 ± 0.05&  5.50& 6.75&   86.69&  53.36& 3.32 ± 0.06\\
            Vector quantization \citep{repcodec}&  4.79&  6.08&   90.05&  51.98& 3.90 ± 0.05&  7.72& 7.56&   86.30&  53.81& 3.50 ± 0.06\\
            \midrule
            WavLM \citep{enhancing-polyglot-voices} &   4.56&   5.84&   89.52&  52.77& 3.84 ± 0.05& 5.14&  6.38&    86.18& 54.30& 3.66 ± 0.06\\
            Proposed (Eta-WavLM) &  \bf 3.81&  \bf 5.63&   \bf 92.46&  \bf 47.60& \bf 4.00 ± 0.05&  \bf 4.64&  \bf 6.09&   \bf 89.32&  \bf 48.25& \bf 3.79 ± 0.05\\
            \bottomrule
        \end{tabular}
   }
\end{center}
\end{table*}
\subsubsection{Experimental Setup}
\label{subsubsec:vc-experimental-results}
To ensure a robust evaluation of the VC system, we selected two English target speakers with distinct background characteristics, genders, and noise levels to train the acoustic model: \textit{LJSpeech} \citep{ljspeech} (F): A single-speaker dataset containing approximately 24 hours of read English speech by a female speaker; \textit{Elliot Miller} (M): A single-speaker dataset consisting of 38 hours of read English speech by a male speaker. We extracted this speaker from the multi-speaker and multi-lingual M-AILABS Speech Dataset\footnote{\url{https://github.com/imdatceleste/m-ailabs-dataset}}. In addition, to ensure a fair comparison with LJSpeech, we randomly selected 24 hours from the dataset. Both target speakers are in the public domain. While LJSpeech is a clean and high-quality dataset, Elliot Miller presents more challenging conditions. This diversity in target speaker profiles was intentionally selected to evaluate the effectiveness of the VC under varied conditions. 

For each audio sample of each target speaker, we first downsample it to 16 kHz and separately extract the corresponding SSL representations using all 6 distinct content encoders. Then, we create the mel-scaled spectrogram of the audio sample following \citep{vocos}, by resampling it to 24 kHz and using the following parameters: $n_{fft} = 1024$, $hop\_length = 256$ and number of Mel bins ($n$-MELs) 100. Finally, for each pair (representation, mel spectrogram), we trained a target-specific acoustic model, using the selected representation as input and the corresponding mel spectrogram as the target. In total, we trained 12 distinct acoustic models (6 types of representations $\times$ 2 speakers). 

\subsubsection{Evaluation Metrics}
\label{subsubsec:vc-evaluation-metrics}
We conducted both objective and subjective evaluations to measure intelligibility, speaker similarity, and overall quality of the converted speech. Intelligibility assesses the system's ability to preserve the linguistic and semantic integrity of the input speech, ensuring that the content is comprehensible after conversion. Speaker similarity evaluates how well the converted speech captures the target speaker's voice characteristics, ensuring the output convincingly mimics the desired speaker. Lastly, overall speech quality examines the naturalness and quality of the converted speech.

To perform these evaluations, we created a test set of 60 utterances obtained by randomly selecting 3 utterances from 20 speakers (10 male and 10 female) extracted from the test-clean set of LibriSpeech. We converted all these utterances into LJSpeech and Elliot Miller using our proposed method and the five baselines, resulting in a total of 420 samples per speaker (60 ground truth + 360 generated). We evaluated \textit{intelligibility} by measuring the word error rate (WER) and phoneme error rate (PER) between the source and converted speech. Orthographic transcriptions were obtained using the Whisper Medium ASR model\footnote{\url{https://huggingface.co/openai/whisper-medium}} \citep{whisper}, while phonetic transcriptions were generated using phonemizer\footnote{\url{https://github.com/bootphon/phonemizer}} \citep{phonemizer}. \textit{Speaker similarity} (SSIM) was measured using a trained speaker verification model\footnote{\url{https://github.com/resemble-ai/Resemblyzer}}. Specifically, we computed the cosine similarity between the d-vectors \citep{d-vector} of each converted sample and those of the source (S-SSIM) and the target (T-SSIM) speakers. Finally, for \textit{overall speech quality}, we conducted a subjective evaluation based on mean opinion scores (MOS). Twenty native-language participants were asked to listen to the randomly mixed samples and rate them on a 5-point scale, where 1 corresponds to ``very poor'' and 5 to ``excellent''.

\subsubsection{Results}
\label{subsubsec:vc-results}
We report the objective and subjective results for the VC experiment. Table \ref{tab:objective-and-subjective-evalution} shows WER/PER, SSIM, and MOS for the two target speakers: LJSpeech (first column) and Elliot Miller (second column). Compared to other methods, Eta-WavLM significantly enhances conversion intelligibility, achieving the lowest error rates for both target speakers. For LJSpeech, it closely approaches the ground truth WER and PER values, demonstrating a high level of linguistic content preservation compared to all baselines. A similar pattern is observed for Elliot Miller, where Eta-WavLM outperforms the baselines, confirming its robustness even under more challenging acoustic conditions. Notably, the perturbation approach exhibits the highest error rates, particularly for Elliot Miller, suggesting that this approach excessively distorts linguistic and semantic information in the input speech. In terms of speaker similarity, the proposed method achieves the best SSIM scores for both target speakers, outperforming both the original WavLM representations and all other disentanglement strategies. While the soft approach yields comparable results for LJSpeech, it struggles with the noisier Elliot Miller speaker, highlighting that some disentanglement methods are more sensitive to challenging acoustic conditions. In contrast, using WavLM directly results in lower speaker similarity, reinforcing the notion that speaker-dependent information remains embedded in the original representations and thus affects the overall performance of the VC system. Finally, regarding overall speech quality, Eta-WavLM achieves the highest MOS scores, surpassing all baselines. Interestingly, the vector quantization approach demonstrates relatively strong MOS ratings but fails to maintain the same level of linguistic and semantic integrity, as evidenced by its higher WER and PER values, especially for Elliot Miller. Conversely, as with intelligibility, the perturbation yields the lowest MOS values, further indicating that speech modification negatively impacts also naturalness. 

These results confirm our hypothesis that Eta-WavLM effectively disentangles speaker information while preserving linguistic content, achieving the best balance between intelligibility, speaker similarity, and speech quality. Moreover, the consistent improvements across both target speakers underline its robustness, demonstrating that the proposed approach not only reduces speaker-related information more effectively than existing methods but also avoids degradation of other features. 

\subsection{Ablation: Speaker Encoder and PCA}
\label{subsec:speaker-encoder-and-pca}
\begin{table}[t]
    \begin{center}
        \begin{adjustbox}{max width=\columnwidth}
            \begin{tabular}{l|ccc|c}
                \toprule
                \multicolumn{1}{l|}{} & \bf WER & \bf PER & \bf T-SSIM & \bf SPK ACC \\
                \midrule
                Resemblyzer w/o PCA&   4.94& 6.01& 89.02& 74.01 ± 0.01 \\
                Resemblyzer w PCA-64&   4.86& 5.92& 89.86& 73.54 ± 0.02 \\
                Resemblyzer w PCA-128&   4.48& 5.84& 90.59& 65.87 ± 0.01 \\
                \midrule
                WavLM-SV w/o PCA&   4.27& 5.81& 89.29& 69.74 ± 0.02 \\
                WavLM-SV w PCA-64&   4.15& 5.75& 89.35& 68.31 ± 0.02 \\
                WavLM-SV w PCA-128&   3.91& 5.70& 89.76& 65.83 ± 0.01 \\
                \midrule
                ECAPA-TDNN w/o PCA&   4.18& 5.80& 89.90& 60.87 ± 0.01 \\
                ECAPA-TDNN w PCA-64&   3.95& 5.63& 90.91& 58.14 ± 0.02 \\
                ECAPA-TDNN w PCA-128&   \bf 3.81& \bf 5.63& \bf 92.46& \bf 55.73 ± 0.01 \\
                \bottomrule
            \end{tabular}
       \end{adjustbox}
       \caption{Results (\%) measuring VC intelligibility (W/PER \(\downarrow\)), target speaker similarity (T-SSIM \(\uparrow\)), and the speaker classification accuracy (SPK ACC \(\downarrow\)) using different speaker encoders and PCA reductions. The VC target speaker is LJSpeech.}
        \label{tab:spk-encoder-and-pca}
    \end{center}
\end{table}
In this section, we analyze the impact of the speaker encoder for the creation of effective speaker embeddings $\mathbf{d}$. Since our approach aims to decompose SSL representations into speaker-dependent and speaker-independent components, it is crucial that $\mathbf{d}$ captures speaker-specific characteristics without encoding other critical components such as linguistic content, prosody, or phonetic details. If the extracted embeddings contain too much non-speaker-related information, the decomposition process of our method risks degrading essential speech content in the SSL representations, resulting in a non optimal \textit{eta} representation $\bm{\eta}$. Furthermore, since embeddings can generally be large and contain redundant information, we also want to investigate whether a technique like PCA to reduce and make more compact $\mathbf{d}$ can further improve the overall performance of our approach. To this end, we evaluated three different speaker encoders: \textit{Resemblyzer}\footnote{\url{https://github.com/CorentinJ/Real-Time-Voice-Cloning}}, a publicly available implementation of \citep{resemblyzer}, released under the MIT license; \textit{WavLM-SV}\footnote{\url{https://github.com/microsoft/UniSpeech/tree/main/downstreams/speaker_verification}}, a WavLM-Large version designed for speaker verification (SV) and released under the Attribution-ShareAlike 3.0 Unported license; and \textit{ECAPA-TDNN} \citep{ecapa-tdnn}, which we briefly introduced in Section \ref{sec:experiments}. For each speaker encoder, we considered the raw output and two levels of dimensionality reduction: PCA-64 and PCA-128. Following the same configurations as in Section \ref{subsec:speaker-related-task} and Section \ref{subsec:voice-conversion}, we setup a Speaker Classification task and a Voice Conversion task (considering only LJSpeech as target speaker), evaluating the \textit{eta} representations created using the different $\mathbf{d}$ generated by each speaker encoder. Table \ref{tab:spk-encoder-and-pca} reports WER/PER and T-SSIM of VC and the mean and standard deviation across the 5 folds of the cross-validation of the speaker accuracy for each model. ECAPA-TDNN consistently achieves the best performance across all metrics, demonstrating its superiority in preserving linguistic content, achieving a high level of target speaker similarity in the VC task, and reaching the lowest speaker classification accuracy. While WavLM-SV also shows strong intelligibility performance, its VC speaker similarity remains lower than that of ECAPA-TDNN. This highlights the fact that the \textit{eta} representations created with $\mathbf{d}$ extracted using WavLM-SV are less speaker-independent than those of ECAPA-TDNN. This is further confirmed by the higher speaker accuracy in the classification task. On the other hand, the performance obtained with Resemblyzer is not comparable to that of the other two approaches, suggesting that the \textit{eta} representations created with its $\mathbf{d}$ are too entangled. Interestingly, reducing the dimensionality of the speaker embeddings using PCA actually enhances overall performance in both the VC and speaker classification tasks for all methods. In this case as well, ECAPA-TDNN achieves the best values across all metrics, particularly with the PCA-128 configuration. This aligns with our hypothesis that reducing redundant information from $\mathbf{d}$ further improves performance. However, excessively reducing the dimensionality of $\mathbf{d}$ does not appear to provide additional benefits. This is evident from the performance obtained using PCA-64, which is lower than that of PCA-128. This suggests that while PCA can enhance performance, its effectiveness depends on the extent of the dimensionality reduction applied. In conclusion, our results demonstrate that ECAPA-TDNN is the most effective speaker encoder for our approach, and applying PCA to $\mathbf{d}$ further enhances the decomposition process, improving intelligibility and preserving essential speech content. 

\section{Conclusion}
\label{sec:conclusion}
In this work, we introduced Eta-WavLM, a novel approach for disentangling speaker-related and speaker-independent components in WavLM representations. By leveraging an innovative decomposition strategy based on a simple linear equation, our method effectively minimizes speaker information while preserving other critical components, such as linguistic content, making it highly suitable for speaker-independent speech processing tasks. We validated its effectiveness through a speaker-related task, confirming its ability to significantly reduce speaker information, and further assessed it on a content-related VC task, demonstrating that Eta-WavLM achieves a superior balance between intelligibility, speaker similarity, and speech quality compared to other existing disentanglement methods. Future work will focus on extending our approach to multilingual settings (including low-resource languages) and integrating our representations into other downstream tasks such as ASR and expressive speech synthesis. Additionally, we plan to explore more sophisticated strategies for disentanglement, including non-linear modeling approaches, to further investigate the potential benefits over our current linear formulation.

\section*{Limitations}
To obtain effective speaker-independent speech representations, we focused on the explicit decomposition of speaker and content components using speaker embeddings. This approach significantly reduces speaker identity leakage, as evidenced by our results showing that \textit{eta} representations created using ECAPA-TDNN yield strong performance. However, our method does not fully eliminate speaker-specific information. In particular, performance in the 10-way speaker classification task remains above chance, suggesting that traces of speaker identity still persist in the resulting features. This residual information may be a consequence of the method’s reliance on the quality of the speaker encoder. Future work could explore alternative speaker representations that further improve the trade-off between content preservation and the removal of speaker-related cues.

Our experiments were conducted using the WavLM model, which has demonstrated state-of-the-art performance in various speech tasks. However, our evaluation primarily focused on English datasets, and the ability to generalize to multilingual speech scenarios remains an open question. We leave to future research the investigation on how well our approach disentangles speaker information while preserving speech content across multiple languages.

We used the LibriSpeech dataset for creating the latent basis $\mathbf{A}^*$ and bias $\mathbf{b}^*$. While LibriSpeech is a large and diverse dataset, we believe that incorporating larger, more diverse datasets, or even multilingual data, could further strengthen the model’s ability to generalize across different linguistic and acoustic environments, ultimately enhancing the robustness and flexibility of our method. This is a direction we plan to pursue in future research.


\bibliography{custom}

\appendix

\section{PaCMAP Visualization}
\label{app:pacmap}
\begin{figure*}[t]
    \centering
    \includegraphics[width=0.90\linewidth]{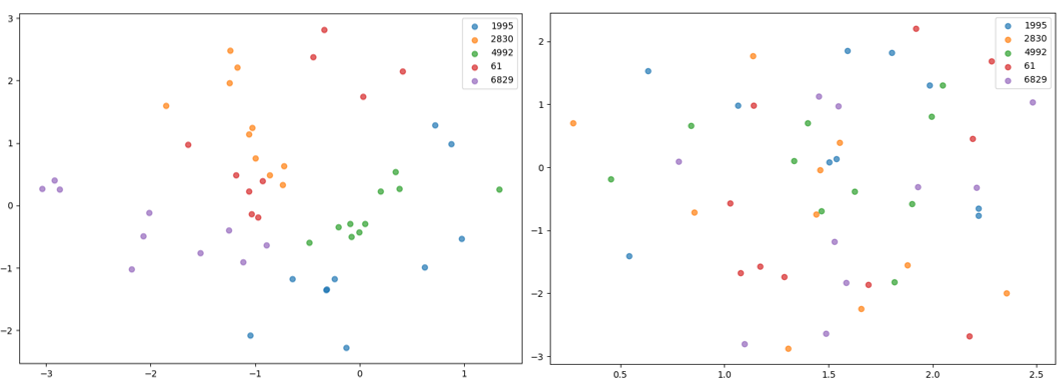}
    \caption{PaCMAP projections of the WavLM (a) and Eta-WavLM (b) representations extracted from 10 utterances of 5 speakers (with ids 1995, 2830, 4992, 61, 6829) from the LibriSpeech test-clean set.}
    \label{fig:pacmap-visualization}
\end{figure*}
In this section, we replicate the analysis from Section \ref{subsec:speaker-related-task} using PaCMAP \citep{pacmap}, an alternative dimensionality reduction technique to UMAP that is known for preserving both global and local data structures. Figure \ref{fig:pacmap-visualization} shows a two-dimensional PaCMAP projection of the same 50 WavLM and Eta-WavLM representations previously visualized using UMAP. In Figure \ref{fig:pacmap-visualization} (a), the WavLM representations exhibit clustering patterns corresponding to individual speakers, once again indicating the presence of speaker-specific information. In contrast, the Eta-WavLM representations in Figure \ref{fig:pacmap-visualization} (b) display no discernible speaker clusters, with utterances more evenly distributed across the space. This additional visualization further supports our findings from the UMAP analysis and provides additional evidence that our transformation significantly reduces speaker-specific information.

\section{Architecture of the Voice Conversion Acoustic Model}
\label{app:vc-am}
In this section, we provide details about the architecture of the acoustic model used for the VC task described in Section \ref{subsec:voice-conversion}, based on \citep{softvc} and \citep{enhancing-polyglot-voices}. The acoustic model takes SSL representations as input rather than graphemes or phonemes as in a typical TTS task and outputs mel spectrograms of the target speaker. The model is composed by an encoder and an autoregressive decoder. Both the encoder and decoder are preceded by a feed-forward pre-net, and a final linear layer with $n$-MELs units follows the decoder. The encoder pre-net is a feed-forward neural network consisting of a stack of two linear layers with 256 units each, ReLU activations, and dropout. The encoder includes a stack of three 1D convolutional layers, each with 512 units, a kernel size of 5, a stride of 1, padding of 2, and ReLU activations. The decoder predicts each spectrogram frame based on the output of the encoder and the previously generated frames. It starts with a decoder pre-net, which is similar in structure to the encoder pre-net, followed by three LSTM layers with 768 units each. Finally, a linear layer with $n$-MELs units generates the output. Furthermore, since there is no attention mechanism between the encoder and decoder, a length regulator module is employed. This module optionally implements a duration adjustment strategy to address potential mismatches between the lengths of the SSL input features and the target spectrogram sequence.

\end{document}